\documentstyle[hip-artc]{article}  

\def\rightmark{Production of resonances} 
\def\leftmark{W. Broniowski, W. Florkowski, B. Hiller}
 
\title{Production of resonances in a thermal model\footnote{\small 
Presented at the {\em 3rd Budapest Winter School on Heavy-Ions}, 8-11~Dec.~2003. }}
 
\authors{
{\twerm  \underline{W.~Broniowski}$^{1}$, W.~Florkowski$^{1,2}$, and B.~Hiller$^3$
}\\[2.812mm]
{\normalsize
\hspace*{-8pt}$^1$ The H. Niewodnicza\'nski Institute of Nuclear Physics, 
Polish Academy of Sciences, PL-31342~Krak\'ow, Poland\\[0.2ex] 
\hspace*{-8pt}$^2$ Institute of Physics, \'Swi\c{e}tokrzyska Academy, PL-25406~Kielce, Poland\\[0.2ex] 
\hspace*{-8pt}$^3$ Centro de F\'{\i}sica T\'eorica, University of Coimbra, P-3004 516 Coimbra, Portugal}}
 
\abstract{We discuss the $\pi^+ \pi^-$ invariant-mass correlations and the 
resonance production in a thermal model with single freeze-out. 
The predictions are confronted with the data from the STAR Collaboration.} 


 
\begin{document}
 
\maketitle

\bigskip

Hadronic resonances, seen in correlation experiments at RHIC, provide valuable
information on the mechanisms of particle production 
in relativistic heavy-ion collisions. 
In this talk we present the $\pi^+\pi^-$ invariant-mass correlations obtained in the
framework of the {\em single-freeze-out model} of Ref.~\cite{sf}. More details 
of this analysis are contained in Ref.~\cite{resonance}.

Here is our {\em single-freeze-out} thermal model in a capsule: 
\begin{enumerate} 
\item The assumption of a single freeze-out, $T_{\rm chem}=T_{\rm kin} \equiv T$, is made.  
\item We incorporate all particles from the Particle Data Tables, with their 
experimental branching ratios.
\item The freeze-out hypersurface is assumed to have the shape similar to the 
Buda-Lund model \cite{BL} (see also \cite{sin}). It possesses a Hubble-like flow, $u^\mu=x^\mu/\tau$.
\item The model has four parameters which depend
on the collision energy. The two thermal parameters, {$T=160$~MeV and $\mu_B=26$~MeV} at
$\sqrt{s_{NN}}=200$~GeV, 
are fixed with the help of ratios of the particle abundances. The geometric 
parameters (the invariant time $\tau$ and the transverse size $\rho_{\rm max}$) 
depend on centrality. They are determined by fitting the 
transverse-momentum spectra \cite{sf}. 
\end{enumerate}

\begin{figure}
\begin{center}
\epsfxsize=9cm \epsfbox{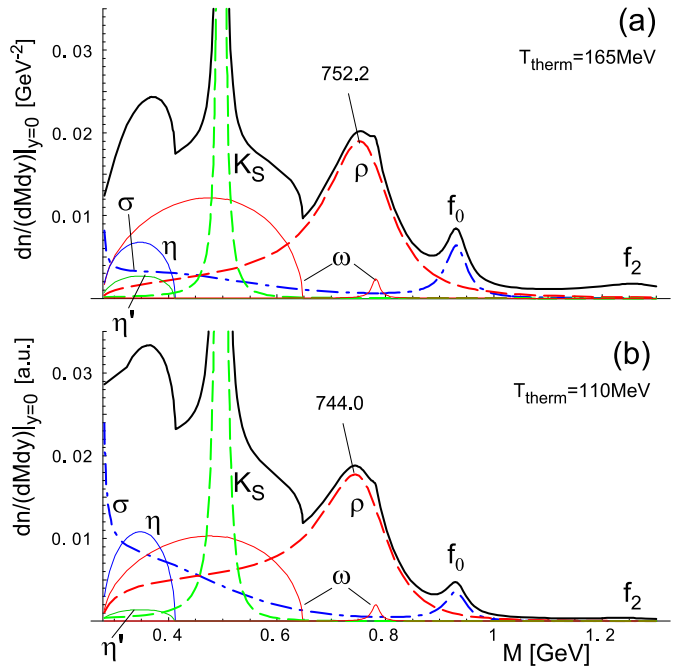}
\end{center}
\vspace{-11mm}
\caption{The $\pi^+ \pi^-$ invariant-mass correlation from decays of 
resonances in a thermal model with a static source. (a) $T_{\rm therm}=T_{\rm chem}=165~{\rm MeV}$.  
(b) $T_{\rm therm}=110~{\rm MeV}$. The numbers indicate the position of
  the $\rho$ peak (in MeV) and the labels indicate the resonance.}
\vspace{-7mm}
\end{figure}

In order to analyze the $\pi^+\pi^-$ invariant mass spectra, measured 
by the STAR collaboration \cite{starrho} with the like-sign subtraction technique, we assume that 
all the correlated pion pairs originate from the decays of neutral resonances.
We include 
$\rho$, $K_S^0$, $\omega$, $\eta$, $\eta'$, $f_0/\sigma$, and $f_2$. 
The phase-shift formula for the volume density of resonances with spin degeneracy
$g$ has the form \cite{phsh} 
\begin{equation}
\frac{dn}{dM} = g \int \frac{d^3 p}{(2\pi)^3} \frac{d {\delta_{\pi \pi}(M)}}{\pi dM} 
\left [ {\exp \left( \frac{\sqrt{M^2+{p}^2}}{T} \right ) - 1} \right ]^{-1}.
\end{equation}
It was used in Ref.~\cite{resonance}; the same formalism has been also applied by
Pratt and Bauer \cite{Pratt} in a similar analysis.

We first show a very simple calculation
where a static source is used. 
Fig.~1 shows the mid-rapidity invariant-mass spectra computed at two
different values of the freeze-out temperature: $T=165$~MeV and $110$~MeV. 
We note a clear appearance of the included resonances, as well as a 
significantly steeper fall-off of the strength in Fig.~1(b), which simply 
reflects the lower temperature.
In that sense the $\pi^+ \pi^-$ invariant-mass spectra provide yet another 
hadronic{\em ``thermometer''} for the temperature at freeze-out.
The collective flow affects the
measured resonance yields due to the presence of kinematic cuts. 
We now show the full calculation in our model, which includes the 
flow, kinematic cuts, and decays of higher resonances.
The complete results are contained in Ref.~\cite{resonance}. 
We have found that the thermal effects and the kinematic cuts 
are able to shift down the position of the $\rho$ peak by about 30~MeV, 
leaving a few tens of MeV unexplained.
For that reason, we have done the calculation with the position of the $\rho$
peak moved down ({\em cf.} Ref.~\cite{theor}) by 9\%.  
In Fig.~2 we compare the model predictions to the STAR data.  
Our model results have been filtered by the detector efficiency 
correction. We note that the model 
does an excellent job in reproducing the gross features of the data. We also note that
the full model, with feeding from higher resonances and with flow and cuts, which
uses the standard single-freeze-out temperature $T=165$~MeV, gives similar predictions
to the naive model at $T=110$~MeV. 
\begin{figure}
\begin{center}
\epsfxsize=8cm \epsfbox{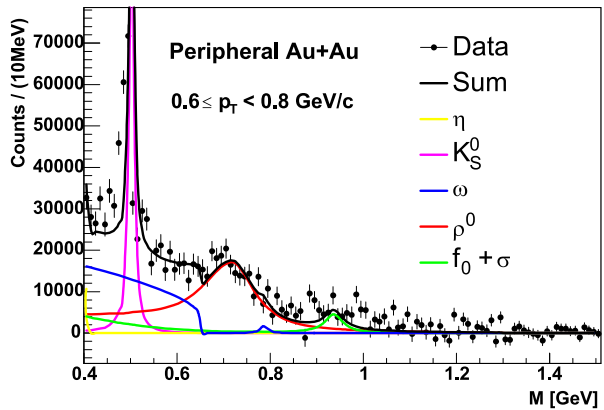}
\end{center}
\vspace{-11mm}
\caption{Single-freeze-out model vs. the data of Ref.~\cite{starrho} for the $\pi^+ \pi^-$ 
invariant-mass correlations. The model calculation includes the 
decays of higher resonances, the flow, the kinematic cuts, and the detector efficiency.}
\vspace{-9mm}
\end{figure}
This is yet another manifestation of the effect of {\em ``cooling''} induced by the resonances
\cite{michal}. Table 1 shows our predictions for yields 
of several resonances for two cases of the $\rho$-meson mass: 770~MeV and 700~MeV.
References to the experimental data can be found in Ref.~\cite{resonance}.

\begin{table}
\begin{center}
{\small
\begin{tabular}{|l|c|c|c|}
\hline
&
$m^*_\rho = 770$ MeV
& 
$m^*_\rho = 700$ MeV 
& Experiment \\ \hline \hline
$T$ [MeV] & $165.6\pm 4.5$ & $167.6\pm 4.6$ & \\ 
$\mu_B$ [MeV] & $28.5\pm 3.7$ & $28.9\pm 3.8$ & \\ \hline \hline
$\eta/\pi^-$ & $0.120 \pm 0.001$  & $0.112 \pm 0.001$ &   \\ 
$\rho^0/\pi^-$ & {$0.114\pm 0.002$} & {$0.135\pm 0.001$} & $0.183 \pm 0.028$ (40-80\%)   \\ 
$\omega/\pi^-$ & $0.108 \pm 0.002$  & $0.102 \pm 0.002$ &   \\ 
$K^*(892)/\pi^-$ & $0.057\pm 0.002$ & $0.054 \pm 0.002$ &    \\ 
$\phi/\pi^-$ & $0.025 \pm 0.001$  & $0.024 \pm 0.001$ &   \\ 
$\eta^\prime/\pi^-$ & $0.0121 \pm 0.0004$  & $0.0115 \pm 0.0003$ &   \\ 
$f_0(980)/\pi^-$ & {$0.0102 \pm 0.0003$} & {$0.0097\pm 0.0003$} & $0.042 \pm 0.021$  (40-80\%)   \\ \hline
$K^\ast(892)/K^-$ & {$0.33\pm 0.01$} & {$0.33\pm 0.01$}
& 
\begin{tabular}{c} 
$0.205 \pm 0.033$  (0-10\%)      \\
$0.219 \pm 0.040$  (10-30\%)     \\
$0.255 \pm 0.046$  (30-50\%)     \\
$0.269 \pm 0.047$  (50-80\%)     \\
\end{tabular} \\ \hline
{$\Lambda(1520)/\Lambda$} & {$0.061\pm 0.002$} & {$0.062\pm 0.002$}
& 
\begin{tabular}{c} 
$0.022 \pm 0.010$  (0-7\%)       \\
$0.025 \pm 0.021$  (40-60\%)     \\
$0.062 \pm 0.027$  (60-80\%)     \\
\end{tabular} \\ \hline
$\Sigma(1385)/\Sigma$ & $0.484\pm 0.004$ & $0.485\pm 0.004$
& $$   \\ \hline
\end{tabular}}
\end{center}
\end{table}

WB and WF are grateful to the organizers of the conference for providing an excellent
scientific and social atmosphere, and to Patricia Fachini for many 
helpful discussions, for filtering our 
results with the STAR detector efficiency, and for preparing Fig.~2. 
This research was supported in part by the Polish State Committee for Scientific Research grant 2~P03B~059~25, 
and by the grants of Funda\c cao para a Ci\^encia e a Technologia, POCTI/FIS/35304/2000 and PRAXIS XXI/BCC/429/94.

\end{document}